\documentclass[notoc]{JINST}
\usepackage{graphicx}
\usepackage{amsmath}
\usepackage{amssymb}
\usepackage{comment}

\def\be{\begin{eqnarray}}
\def\ee{\end{eqnarray}}
\def\bea{\begin{eqnarray}}
\def\eea{\end{eqnarray}}
\def\beas{\begin{eqnarray*}}
\def\eeas{\end{eqnarray*}}

\def\nn{\nonumber}
\def\ve{\varepsilon}

\title{A concept for the experimental determination of the nucleon electric to magnetic form factor ratio
at very low $Q^2$}
\author{G.~Ron$^a$\thanks{Corresponding author.}, E.~Piasetzky$^b$, and B.~Wojtsekhowski$^c$\\
\llap{$^a$}{Department of Particle Physics, Weizmann Institute of Science, Rehovot 76100, Israel (\email{guy.ron@weizmann.ac.il})}\\
\llap{$^b$}{The Beverly and Raymond Sackler School of Exact Sciences,
Tel Aviv University, Tel Aviv 69978, Israel}\\
\llap{$^c$}{Thomas Jefferson National Accelerator Facility, Newport News, VA 23606, USA}}

\abstract{
Stationary target measurements of the nucleon form factors have been performed with high precision down to $Q^2$ 
of $\sim$ 0.01 GeV$^2$ for protons ($G_E^p$) and down to
$\sim$ 0.1 GeV$^2$ for neutrons ($G_M^n$). Conventional extraction using cross section and 
polarization measurement
cannot be extended to very low values of $Q^2$ due to inherent experimental limitations.
We present a proposal for a new approach to a measurement, using colliding beams, which will 
extend the range of possible measurement at low $Q^2$ by several orders of magnitude 
over stationary target limits.
}

\begin{document}

\section{Introduction}
The Dirac and Pauli form factors of the nucleon, $F_1$ and $F_2$, commonly expressed as the Sachs electric
and magnetic form factors~\cite{Sachs62}
\bea
\nn G_E&=&F_1-\tau F_2\\
G_M&=&F_1+F_2
\eea
where $\tau=Q^2/4M_N^2$ and $M_N$ is the nucleon mass, are fundumental observables describing the electromagnetic current
of the nucleon, 
{\flushleft
\be
\nn\langle p',\lambda'| J^\mu(0)| p,\lambda\rangle
&=&\bar{u}(p',\lambda')\bigg(\gamma^\mu F_1(Q^2) \left.+i{\sigma^{\mu\alpha}\over 2M}q_\alpha F_2(Q^2)
\right) u(p,\lambda).
\ee
}
For a recent review of the nucleon form factors see~\cite{Perd07} and references therein.

In the Breit frame the electric (magnetic) form factor is related to the charge (magnetization)
density distribution by a Fourier transform,
\be
G_{E(M)}(Q^2)=\int \rho_{Ch(M)}(\vec{r})e^{-i\vec{q}\cdot\vec{r}}d^3r,
\ee
where $\vec{q}$ is the three momentum transfer and $Q^2=-q^2=\vec{q}^2-\omega^2$, and 
$\omega$ is the energy transfer ($\omega=E-E'$). Using this definition
one may extract the Breit frame RMS charge and magnetization radii via a low $Q^2$ expansion of the 
form factors,
\be
\left<r^2\right>=-\left.6\frac{d G(Q^2)}{dQ^2}\right|_{Q^2=0},
\ee
or, equivalently, by fitting the slope of the form factor at very low $Q^2$. While this extraction 
is strictly only valid in the non-relativistic regime and in the Breit frame it has been shown~\cite{Miller1,Miller2}
that by using light-front formalism it is possible to relate the Dirac and Pauli form factors to the 
{\it{transverse}} Dirac and Pauli radii via a two dimensional Fourier transform,
\be
\label{eq:impact}
F_{1(2)}(Q^2)=\int\rho(\vec{b})_{D(P)}e^{-i\vec{q}\cdot\vec{b}}d^2b,
\ee
where $\vec{b}$ is the transverse radius vector. Note that eq. (\ref{eq:impact}) is
valid relativistically and in all reference frames. It was also shown~\cite{Miller2} that
it is possible to relate the difference of the RMS transverse Dirac and Pauli radii,
\be
\label{eq:LC_Approx}
\nn F_1(Q^2)&\approx&1-\frac{Q^2}{4}\left<b^2\right>_{D},\\
F_2(Q^2)&\approx& \kappa\left(1-\frac{Q^2}{4}\left<b^2\right>_P\right),
\ee
where $\left< b^2\right>$ is the RMS transverse radius, and the relation is valid for low $Q^2$. Eq. (\ref{eq:LC_Approx}) was used 
to show that the magnetization density of the proton extends further than the charge density, however,
the paucity of the data at low $Q^2$ is detrimental to an accurate determination of the difference.

The low Q2 property of the nucleon is a subject of chiral QCD~\cite{Meissner}.
It is of great interest for to discover at what $Q^2$ predictions of chiral QCD
are in agreement with the nature.

We further note that the values of the form factors at low $Q^2$ effect several other 
experimental results and calculations, such as, for example, calculations of the Zemach radius~\cite{Zemach},
the extraction of Generalize Parton Distributions (GPDs)~\cite{GPDS} from experiment (via the
Bethe-Heitler interference terms), and the extraction of strange form factors 
from parity violation measurements~\cite{PV}.

We conclude that accurate measurements of the nucleon form factors at low $Q^2$ are desirable, as evident
in the approval of several recent experiments~\cite{E08007,Mainz} targeting the low $Q^2$ energy 
range.

\section{Form Factor Measurements}
The cross section for elastic scattering of an electron off the nucleon may be written as~\cite{Rosenbluth}:
\be
\label{eq:xs}
\frac{d\sigma}{d\Omega_e}&=&\left(\frac{d\sigma}{d\Omega}\right)_{Mott}\frac{E'}{E}\left\{F_1^2(Q^2)+2(F_1(Q^2)+F_2(Q^2))^2 \tan^2\frac{\theta_e}{2}\right\}\\
\nn&=&\left(\frac{d\sigma}{d\Omega}\right)_{Mott}\frac{1}{1+\tau}\left(G_E(Q^2)^2+\frac{\tau}{\varepsilon}G_M(Q^2)^2\right),
\ee
where 
\be
\label{eq:Mott_xs}
\left(\frac{d\sigma}{d\Omega}\right)_{Mott}=\left(\frac{e^2}{2E}\right)^2\left(\frac{\cos^2\frac{\theta_e}{2}}{\sin^4\frac{\theta_e}{2}}\right),
\ee
is the Mott cross section for the scattering of a spin 1/2 electron from a spin-less, point-like target, 
$E/E'=(1+2E/M_p\,\sin^2\theta_e/2)^{-1}$ is the recoil factor, 
$\theta_e$ is the scattered electron angle, and $\ve\equiv\left[1+2(1+\tau)\tan^2(\theta_e/2)\right]^{-1}$ 
is the polarization of the virtual photon. 

The traditional method of extracting the EM form factors is the so called "Rosenbluth separation" method~\cite{Rosenbluth} in which
the EM cross section is measured for the same value of $Q^2$ and different values of $\ve$ and a linear fit is
used to determine the form factors. While this method has provided valuable data over the years, there are intrinsic limitations
to the technique, easily discerned by examination of eq. (\ref{eq:Mott_xs}). At high $Q^2$, the cross section 
is dominated by the term proportional to $G_M^2$, making an extraction of the $G_E$ term increasingly difficult. 
Conversely, at low $Q^2$ the $G_M$ term is suppressed by $Q^2$. Some experiments have measured the cross section at 
$\theta_e\sim180^\circ$, where $\ve=0$ and the cross section contains only the $G_M$ term, while useful, these measurements
must take into account the finite detector acceptance which allows contamination of the measurement from the $G_E$ term.

\begin{figure}[ht]
\begin{center}

\includegraphics[width=0.9\textwidth]{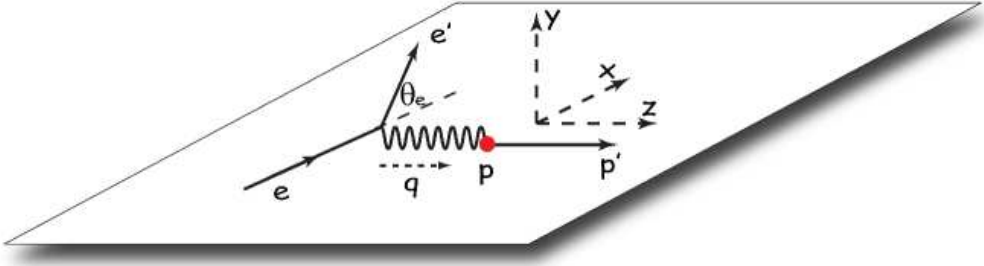}
\end{center}
\caption{\label{fig:PolScat}A schematic illustration of the recoil polarimetry method.}
\end{figure}

In the last two decades an alternative method has been developed and employed. Using the 
high current, high polarization beams and highly polarized targets which have become available. 
In a {\it recoil polarization} measurement a longitudinally polarized electron beam is elastically scattered off an unpolarized nucleon and 
the polarization of the outgoing nucleon is measured. The polarization components of the outgoing nucleon (in 
the Born approximation) are~\cite{Akh74}:
\be\nn \sigma_{red}C_x&=&-2h\,\cot\frac{\theta_e}{2}\sqrt{\frac{\tau}{1+\tau}}G_E G_M,\\
\sigma_{red}C_z&=&h\frac{E+E'}{M_N}\sqrt\frac{\tau}{1+\tau}G^2_M,
\ee
where $C_z$ is the longitudinal polarization, $C_x$ is the transverse, in plane, polarization of the scattered
nucleon, $\sigma_{red}$
is the reduced cross section, 
\be
\label{eq:xs_red}
\sigma_{red}=(1+\tau)\frac{d\sigma/d\Omega}{(d\sigma/d\Omega)_{Mott}}=G_E^2+\frac{\tau}{\ve}G_M^2,
\ee
$h$ is the beam helicity, and all other quantities are defined in (\ref{eq:xs}). Note
that in the Born approximation the induced polarization, normal to
the scattering plane, vanishes. Thus, the ratio of the electric to the magnetic
form factors can be easily determined as:
\be
\mu_N\frac{G_E}{G_M}&=&-\mu_N\frac{C_x}{C_z}\frac{E+E'}{2M_N}\tan\frac{\theta_e}{2}
\ee
Figure \ref{fig:PolScat} presents an illustration of the recoil polarimetry method.

\begin{figure}[ht]
\centering
\includegraphics[angle=0,width=0.8\textwidth]{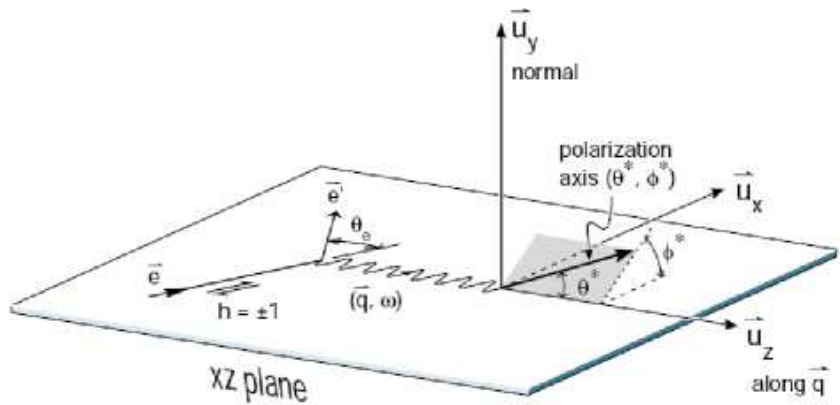}
\caption{\label{fig:PolPolScat}An illustration of the the reaction $\vec{p}(\vec{e},e')p$.}
\end{figure}

In a {\it Beam Target Asymmetry} measurement a longitudinally polarized electron beam is scattered off a polarized target and the 
asymmetry of the scattering cross section is measured. The asymmetry may be written as (neglecting the target dilution factor):

\be
A&\equiv&\frac{\sigma_+-\sigma_-}{(\sigma_++\sigma_-)}\\
&=&-\frac{2\sqrt\frac{\tau}{1+\tau}\tan\frac{\theta}{2}\left\{
\sqrt{\tau\left(1+\left(1+\tau\right)\tan^2\frac{\theta}{2}\right)}\cos\theta^* G_
M^2+\sin\theta^*\cos\phi^* G_M G_E\right\}}
{\left(\frac{G_e^2+\tau G_M^2}{1+\tau}+2\tau G_M^2 \tan^2(\theta/2)\right)},\nonumber
\ee
where $\theta^*$ ($\phi^*$) is the polar (azimuthal) angle of the momentum vector of the recoil nucleon with respect
to the target polarization vector. Figure \ref{fig:PolPolScat} illustrates the beam target asymmetry 
method.

While spin correlation measurements have enabled the extension of the measured range of the form factor ratio to $Q^2$ as
low as 0.01 GeV$^2$~\cite{E08007} there exist fundamental difficulties in extending the range even further which may not be overcome in 
fixed-target or in-beam 
target experiments. 

In a fixed target experiment the energy of the recoil nucleon in elastic scattering is directly proportional to the 
value of $Q^2$, namely,
\be
T_N=\omega=\frac{Q^2}{2M_N},
\ee
where $T_N$ is the kinetic energy of the recoil nucleon.
Thus, for low $Q^2$ the nucleon is ejected with very low kinetic energy, severely restricting the possibility of detecting 
it (even before requiring a secondary scattering reaction). One might suppose that by using beam-target asymmetry measurements
it is possible to forgo detection of the recoil nucleon and detect instead the scattered electron. However, for low $Q^2$, the
electron is ejected with very forward angle, again restricting the applicability of this method (cf. for example~\cite{E08007}). 

\section{A new proposal}
Here we present a proposal for an alternative technique, based on the recoil polarization method, which
will enable form factor ratio measurements to be extended to extremely low values of $Q^2$. The
proposed measurement is based on the fact that $Q^2$ is a Lorentz invariant quantity. 
We propose
to use two colliding beams, of unpolarized nucleons and of longitudinally polarized electrons, such that for 
low values of $Q^2$ the colliding particles retain almost all of their original momentum (in the lab frame), and are 
easy to detect. The polarization of the scattered nucleon may then be measured using a polarimeter, and the polarization 
of the nucleon in the rest frame may be calculated using a Lorentz transformation (which, as stated previously, 
leaves $Q^2$ unchanged). An alternative method, with the advent of polarized storage rings, is to 
collide two polarized beams and detect the scattered electron (analogous to the beam target asymmetry 
measurement).

An idea of the strength of this technique may be gained by examining some representative 
kinematic setting for a 500 MeV electron beam and a 40 MeV (kinetic energy) proton beam, which are summarized in Table~\ref{tab:esrkin}
(where the initial electron beam angle is 0$^\circ$ and that of the proton beam is 180$^\circ$). Clearly, the scattered particles
may be detected, even for very low value of $Q^2$.

\begin{table}[ht]
\begin{center}
\begin{tabular}{|c|c|c|c|c|}\hline
$Q^2$&  {$\theta_e$}& $E'_e$& $\theta_p$& $T_{p'}$ \\ 
(GeV$^2$)& {(deg)}& (GeV)& (deg)& (GeV) \\ \hline
0.55& 111.55& 0.403& 134.75& 0.138\\ \hline
0.1& 37.56& 0.482& 61.58& 0.058 \\ \hline 
0.01& 11.5& 0.498& 20.53& 0.042 \\ \hline
0.001& 3.62& 0.499& 6.54& 0.0402 \\ \hline
0.0006& 2.8& 0.499& 5.066& 0.0401 \\ \hline
0.0001& 1.14& 0.499& 2.07& 0.040 \\ \hline
\end{tabular}
\caption{\label{tab:esrkin}Representative kinematics for a 500 MeV electron beam and a 40 MeV proton beam.}
\end{center}
\end{table}

\begin{figure}[ht]
\centering
\includegraphics[angle=0,width=0.8\textwidth]{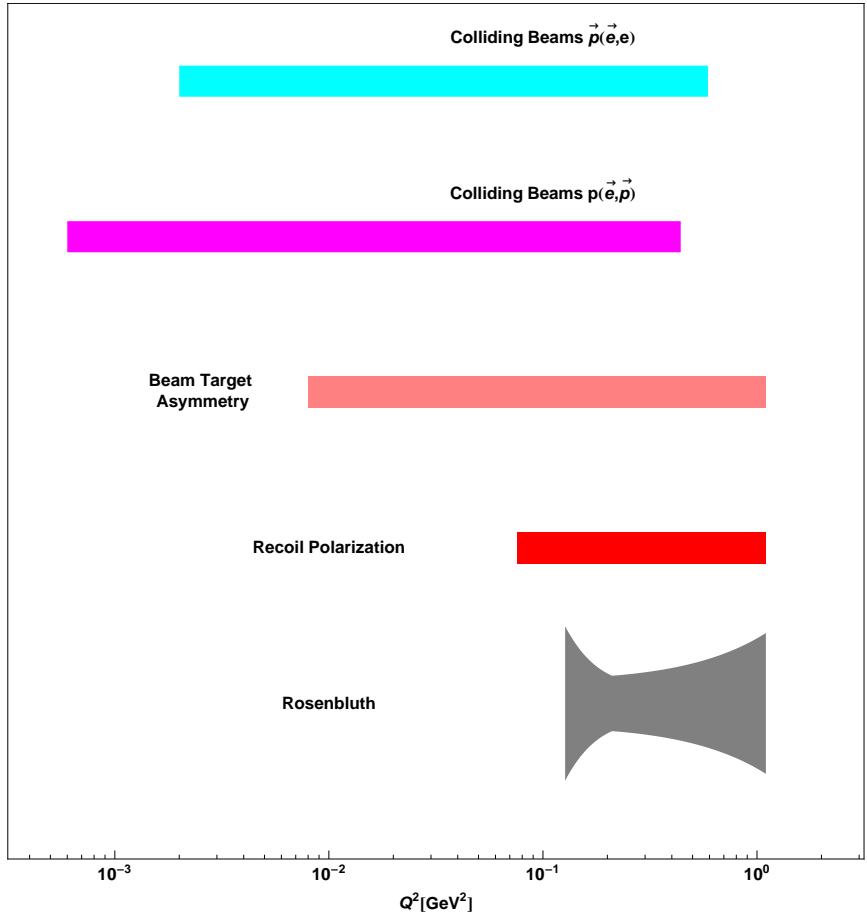}
\caption{\label{fig:cover}An illustration of the the ranges of applicability and the uncertainties for the
different methods.}
\end{figure}

An illustrative representation of the $Q^2$ range of the different experimental techniques 
is presented in Figure \ref{fig:cover},
the possible relative (statistical) uncertainties
on the form factor ratio, $\Delta R/R$ are also shown. In order to compare the
different methods we use the following assumptions:
\begin{itemize}
\item Scattered particles may be detected at angles of 5$^\circ$ to 120$^\circ$.
\item The beam energy varies between 200 MeV and 4.4 GeV.
\item Secondary scattering in a polarimeter is possible for protons whose kinetic energy exceed 40 MeV.
\item In order to remove the dependence on beam time allocation we assume 
that enough beam time is allocated to measure the cross section with 1\% uncertainties and the 
polarizations with the same uncertainty (due to the large cross section at low $Q^2$ this is
a valid assumption, cf. \cite{E08007,Mainz}).
\item For the colliding beams we assume a 500 MeV electron beam colliding with a 40 MeV proton beam.
\item For clarity we omit regions of $Q^2$ where the statistical uncertainty is larger than 5\%.
\end{itemize}

Table \ref{tab:restricts} gives an overview of the restrictions on the different methods. For each 
method in Fig. \ref{fig:cover} the high and low $Q^2$ cutoff values are explained.

\begin{table}[ht]
\begin{center}
\begin{tabular}{|l|c|c|}\hline
Method& Low $Q^2$ Cutoff& High $Q^2$ Cutoff\\ \hline
Collider $\vec{e}(\vec{p},e')$& $\theta_{e'}$ $<$ 5$^\circ$& $\theta_{e'}$ $>$ 120$^\circ$\\ \hline
Collider $\vec{e}({p},\vec{p'})$& $\theta_{p'}$ $<$ 5$^\circ$& $\theta_{p'}$ $>$ 120$^\circ$\\ \hline
Beam target asymmetry& $\theta_{e'}$ $<$ 5$^\circ$& -\\ \hline
Recoil polarization& $T_{p'}$ $<$ 40 MeV& -\\ \hline
Rosenbluth separation& $\Delta R/R$ $>$ 5\%& $\Delta R/R$ $>$ 5\%\\\hline
\end{tabular}
\caption{\label{tab:restricts}Restricting parameters for the high and low $Q^2$ cutoff values of each method.}
\end{center}
\end{table}

\begin{figure}[ht]
\centering
\includegraphics[angle=0,width=.65\textwidth]{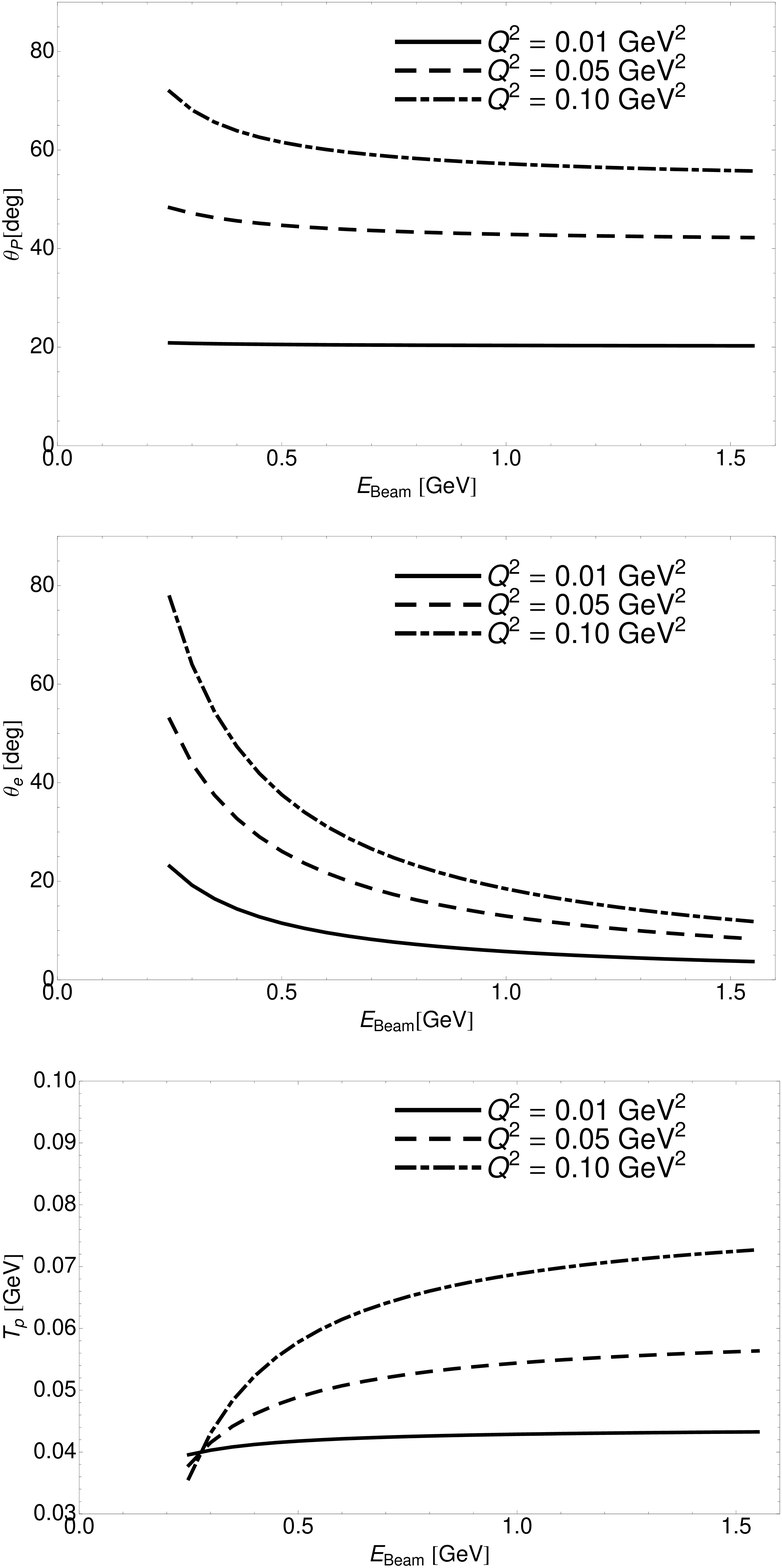}
\caption{\label{fig:Col}$\theta_e$, $\theta_p$ and $T_p$ vs. electron beam energy
for several different $Q^2$ values.}
\end{figure}

Figure \ref{fig:Col} shows several $Q^2$ angle vs. electron beam energy and $T_p$ vs. beam energy
contours for a proton beam of 40 MeV energy 
colliding with an electron beam of energy between 100 MeV and 1.5 GeV.

In order to present a quantitative idea about the applicability of this method we calculate the
luminosity for a possible setup (electron
storage ring and  proton beam parameters are taken from the design 
specs for the ELISe ring~\cite{ELISe} and the SARAF accelerator~\cite{SARAF}, respectively).
Table \ref{tab:params} lists the parameters used for the electron 
storage ring and the linear proton beam. The calculated luminosity is ${\cal L}$ $\sim$ $10^{29}$ cm$^{-2}$sec$^{-1}$.
Using these values for $Q^2$ = 0.0005 $GeV^2$  the calculated cross section is $d\sigma/d\Omega$ $\sim$ 5.3$\cdot$10$^{-26}$
 cm$^{-2}$ which translates into an event rate of ${\cal O}(10 Hz)$ for a few mrad detector acceptance.

\begin{table}[ht]
\begin{center}

\begin{tabular}{|ll|}\hline
Electron Storage Ring &\\
Circumference (m)& 45.215 \\ 
Energy (MeV)& 250 \\ 
Number of Bunches& 8\\
Revolution Freq. (MHz)& 6.63\\
Bunch Population& 5$\cdot$10$^{10}$\\
Current (mA)& 425\\
Beam Emittance, $\ve_{x,y}$ ($\mu$m$\cdot$mrad)& 50\\ 
Beam Size at IP, $\sigma_{x,z}$ ($\mu$m)& 220\\
Bunch Length (cm)& 4\\
Total RF Generator Power (kW)& 42\\\hline
Proton beam & \\
Energy (MeV)& 40\\
Current (mA)& 2\\
Bunch Frequency (MHz)& 176\\
Bunch Population& 7$\cdot$10$^{10}$\\
Beam Radius at IP (mm)& 0.6\\\hline
\end{tabular}
\caption{\label{tab:params}Parameters used for the luminosity estimation~\cite{ELISe,SARAF}.}
\end{center}
\end{table}
\subsection*{Extensions}
Further intriguing extensions of this method are possible, here we note a few of the more interesting ones:
\begin{itemize}
\item A storage ring may be designed to accommodate both electron and positron beam and 
both proton and anti-proton beams are either available or will be in the near future. Thus, a comparison may be made
of the measured asymmetries for the reactions:
\be
\nn e^-+p&\to&e^-+p\\
\nn e^++p&\to& e^+ + p\\
\nn e^-+\bar{p}&\to&e^-+\bar{p}\\
e^++\bar{p}&\to&e^++\bar{p}
\ee
which may be used as a test for the equivalence under C parity of the proton and anti-proton.
\item Availability of a polarized proton beam (such as the one planned for the FAIR facility at GSI/Darmstadt\cite{FAIR})
will remove the need for the measurement of the outgoing proton polarization, requiring instead the (perhaps 
easier) measurement of the asymmetry in the $\vec{p}(\vec{e},e')$ or $\vec{p}(\vec{e},p')$ reaction.
\end{itemize}
\subsection*{Detection Scheme}
In the simple 2-body reaction envisioned here detection of one of the final particles determines the kinematics 
of the reaction. 
The proposed approach for measurement of the form factors allows 
the proton to have significant energy after the scattering, 
even for very small momentum transfer, drastically simplifying its detection.

When the proton beam is polarized, the final proton could be detected, for
example, by a semi-conductor detector with very good energy
and position resolutions, leading to simple selection of the elastic
scattering events.

Detection of scattered electrons with good energy resolution is possible
by making use of the storage ring magnet to perform momentum analysis (for
small scattering angles) or an ordinary magnetic spectrometer for larger angles.

For proton (antiproton) energies of several tens of MeV
polarimetery via secondary scattering becomes an attractive option as
the figure of merit for such polarimeters can be as high as 1\% (see for example~\cite{Glister:2009cu} and references therein).
Since the luminosity of the proposed experiments will be on the level of 
$10^{30}-10^{32}$ cm$^{-2}$s$^{-1}$ we expect no problems with the operation of 
segmented detectors even at small angles.

\section{Summary}
In summary, we have presented a method by which the nucleon electric to magnetic form factor ratio may be 
measured to high precision and low momentum transfer. Such a measurement will allow an accurate determination 
of the difference of the nucleon electric and magnetic radii, as well as impact many atomic high precision 
experiments. The experimental requirements to perform such a measurement are well understood and several possible 
facilities exist which may be able, with slight modifications, to perform such a measurement. 

We thank R.~Gilman for many illuminating discussions. We thank
Y.~Hammer for the kinematical calculations. We also thank the Israeli Science Foundation for partial support 
of this work. This work
was partially supported by DOE contract DE-AC05-84ER40150
under which the Southeastern Universities Research Association
(SURA) operated the Thomas Jefferson National
Accelerator Facility.

\end{document}